\documentclass{jpsj2}
\title{$f$-Electron-Nuclear Hyperfine-Coupled Multiplets in the Unconventional Charge Order Phase of Filled Skutterudite PrRu$_4$P$_{12}$} 
\author{Yuji \textsc{Aoki}$^{1}$ \thanks{E-mail address: aoki@tmu.ac.jp}, Takahiro \textsc{Namiki}$^{1}$, Shanta R.~\textsc{Saha}$^{1}$ \thanks{Present address: Center for Nano Physics and Advanced Material, Department of Physics, University of Maryland, MD 20742, USA} , Takashi \textsc{Tayama}$^{2}$, Toshiro \textsc{Sakakibara}$^{3}$, Ryousuke \textsc{Shiina}$^{4}$, Hiroyuki \textsc{Shiba}$^{5}$, Hitoshi~\textsc{Sugawara}$^{6}$ and Hideyuki \textsc{Sato}$^{1}$}
\inst{$^{1}$Department of Physics, 
Tokyo Metropolitan University, Tokyo 192-0397, Japan \\
$^{2}$Department of Physics, 
Toyama University, 3190 Gofuku, Toyama 930-8555, Japan \\
$^{3}$Institute for Solid State Physics, 
University of Tokyo, Kashiwa, Chiba 277-8581, Japan \\
$^{4}$Department of Materials Science and Technology, 
Niigata University, Niigata 950-2181 \\
$^{5}$Department of Physics, 
Tokyo Institute of Technology, Tokyo 152-8551, Japan \\
$^{6}$Department of Physics, Kobe University, Kobe 657-8501, Japan
}

\abst{
The filled skutterudite PrRu$_4$P$_{12}$ is known to undergo 
an unconventional charge order phase transition at 63 K, below which 
two sublattices with distinct $f$-electron crystalline-electric-field 
ground states are formed. In this paper, we study experimentally 
and theoretically the properties of the charge order phase at very 
low temperature, particularly focusing on the nature of the degenerate 
triplet ground state on one of the sublattices. 
First, we present experimental results of specific heat and 
magnetization measured with high quality single crystals. 
In spite of the absence of any symmetry breaking, the specific heat 
shows a peak structure at $T_{\rm p}=0.30$ K 
in zero field; it shifts to higher temperatures 
as the magnetic field is applied. In addition, the magnetization curve has a remarkable rounding below 1 T. 
Then, we study the origin of these experimental findings by considering 
the hyperfine interaction between 4$f$ electron and nuclear spin. 
We demonstrate that the puzzling behaviors at low temperatures can be 
well accounted for by the formation of {\it 4$f$-electron-nuclear 
hyperfine-coupled multiplets}, the first thermodynamical 
observation of its kind. 
}

\kword{4$f$-electron-nuclear hyperfine-coupled multiplets, PrRu$_4$P$_{12}$, charge ordering, filled skutterudite, crystalline electric field} % \ldots

\begin{document}
\maketitle

\section{Introduction}

Pr-based filled skutterudites PrT$_4$X$_{12}$ have received intense 
interest since they display a variety of exotic strongly-correlated-electron behaviors~\cite{AokiK40JPSJ2005, skut2008, SatoMM, BauerPOSPRB2002,AokiJPSJ2005, AokiPFPPRB2002,KikuchiJPSJ2007}.
Among them, a charge-ordering transition in PrRu$_4$P$_{12}$ 
appearing at $T_{\rm co}=63$ K from a high-$T$ metallic phase 
to a low-$T$ nonmetallic phase is unique~\cite{SekinePRL1997}. 
No clear anomaly at $T_{\rm co}$ in the magnetic 
susceptibility suggests that this ordering has a nonmagnetic origin. 
Above $T_{\rm co}$, the crystal structure is a body-centered cubic 
({\it bcc}) with the space group Im$\overline{3}$ ($T_h^5$, \#204).
Below $T_{\rm co}$, the unit cell is doubled in size and the resulting 
simple cubic ({\it sc}) lattice (Pm$\overline{3}$, $T_h^1$, \#200) 
contains two crystallographically-inequivalent Pr sites 
(referred to as Pr1 and Pr2 hereafter)~\cite{LeePRB2004}. 
Both the unit-cell doubling and the decrease of the carrier density 
in the ordered state can be accounted for as being due to 
the Fermi-surface nesting with a wave vector 
${\mib q}=(1, 0, 0)(2\pi/a)$ of the conduction band consisting 
mainly of the $a_u$ molecular orbital on 
P$_{12}$~\cite{HarimaJPCM2003,HarimaJPSJ2007}. 
However, the charge ordering in PrRu$_4$P$_{12}$ is unconventional in the sense that the 4$f$-electrons of Pr ions play an essential 
role in the ordering, since such a transition is absent in 
LaRu$_4$P$_{12}$~\cite{Saha2005}. 
This is also supported by a significant $T$ dependence of the 
crystalline-electric-field (CEF) level schemes of the two Pr 
sites~\cite{IwasaPRB2005,IwasaJPSJ2005,OgitaRaman2009}, 
resulting in a $\Gamma_1$ singlet ground state at the Pr1 site 
and a $\Gamma_4^{(2)}$ triplet ground state at the Pr2 site 
for $T \ll T_{\rm co}$.

Some theoretical studies have been made to understand 
the nature of the charge ordering and the 
metal-non-metal transition. 
The characteristic behavior of the $T$ dependent 
CEF level schemes has been explained as a multipole ordering 
of 4$f$ electrons associated with charge ordering of 
conduction electrons~\cite{TakimotoJPSJ2006}. 
Anomalous $T$ dependences of the electrical resistivity and 
Hall coefficient (with a sign change in the latter) below 
$T_{\rm co}$~\cite{Saha2009} have been explained by taking into 
account 4$f$-orbital fluctuations~\cite{ShiinaJPSJ2009}.
A recent microscopic derivation of an effective Hamiltonian 
has shown that the charge order 
originates simply and exclusively from the $c$-$f$ mixing whose magnitude 
depends strongly on the CEF states.~\cite{ShiinaShibaJPSJ2010} 
Concerning the nature of the triplet ground state in the charge 
order phase, the effective Hamiltonian predicts that the ratio 
of the possible magnetic transition 
temperature to $T_{\rm co}$ should be unusually small, that is, 
of the order of $10^{-3}$.\cite{ShiinaShibaJPSJ2010} 
In fact, no sign of long-range order 
has been detected so far below $T_{\rm co}=63$ K 
in PrRu$_4$P$_{12}$ (no magnetic ordering down to 20 mK by 
zero-field $\mu$SR~\cite{SahaPB2005}). Thus, one important 
issue to be clarified is whether and how the degeneracy 
of the triplet CEF ground state is lifted as $T$ approaches $T=0$. 

For this purpose, we report in the present paper our results of specific heat 
and magnetization measurements in the temperature interval 
$T=0.1 \sim 10$ K on PrRu$_4$P$_{12}$ single crystals. 
It is shown in these experiments that a distinct peak structure 
appears in the specific heat at around $T_{\rm p}=0.30$ K, 
which shifts to higher temperatures as the magnetic field is applied. 
In addition, it is revealed that the magnetization curve displays 
a remarkable rounding below 1 T. Note again that these anomalies 
cannot be attributed to any phase transition.  Remembering that the Pr ion has a large hyperfine coupling,  we consider the effect of the on-site hyperfine interaction 
between the triplet magnetic moment and the $^{141}$Pr nuclear 
spin $I=5/2$. Then we find that the experimental results can be  
reproduced even quantitatively as a  formation of new multiplets 
due to the hyperfine interaction (hereafter called 
{\it 4$f$-electron-nuclear hyperfine-coupled multiplets}).

It is to be noted that although the hyperfine coupling strength 
is relatively enhanced in rare-earth ions, 
its energy scale ({\it e.g.}, $A=+0.052$ K for Pr 
ion~\cite{KondoHC1961,BleaneyHC1963,AokiPFPPRB2002}) is usually 
still smaller than those of inter-ion exchange interactions 
of various types (dipolar, quadrupolar, etc.).
Therefore, most Pr-based compounds having a triplet 
ground state order magnetically at low temperatures, preventing 
such phenomenon. In this sense, the formation of the lattice of 
{\it 4$f$-electron-nuclear hyperfine-coupled multiplets} 
without showing any long-range orderings in PrRu$_4$P$_{12}$ 
is an extremely rare case.
To the best of our knowledge, this is the first 
thermodynamical observation of its kind.

This paper is organized as follows. After a description of 
the experimental details in the next section, 
we present the results of specific heat and magnetization 
measurements in \S 3. We compare them with the theoretical 
results based on the hyperfine interaction in \S 4, and 
discuss some remaining problems in \S5. 
The final section (\S6) is devoted to the summary of the paper. 
In the Appendix, we describe in detail the pseudo-spin 
representation for the ground triplet in a Pr ion 
to explain the new multiplet formation.

\section{Experimental Details} % --------------
Two single crystals, each used for the specific heat and magnetization measurements, are from the same batch of PrRu$_4$P$_{12}$ grown by the Sn-flux method 
using high-purity raw materials of 4N(99.99\% pure)-Pr, 4N-Ru, 6N-P 
and 5N-Sn. The single crystalline nature has been checked by X-ray back-reflection Laue technique. No impurity phases were detected 
in an X-ray powder diffraction experiment for crystals from the same batch. 
Observed de Haas-van 
Alphen oscillations in LaRu$_4$P$_{12}$ single crystals grown 
by the same manner~\cite{Saha2005} attest to high quality of the 
present samples. Specific heat was measured by a quasi-adiabatic 
heat pulse method using a $^3$He-$^4$He dilution refrigerator equipped 
with an 8 T superconducting magnet. For the measurement of dc 
magnetization, a capacitive Faraday force magnetometer 
installed in a dilution refrigerator~\cite{Sakakibara1994} 
was used below 2 K and a SQUID magnetometer (MPMS, Quantum Design Co.) 
above 2 K. In both measurements, the magnetic field was 
applied along  [100], [110], 
or  [111] direction.

\section{Results} % -------
\begin{fullfigure}[t] % 
\begin{center}
\includegraphics[width=13cm]{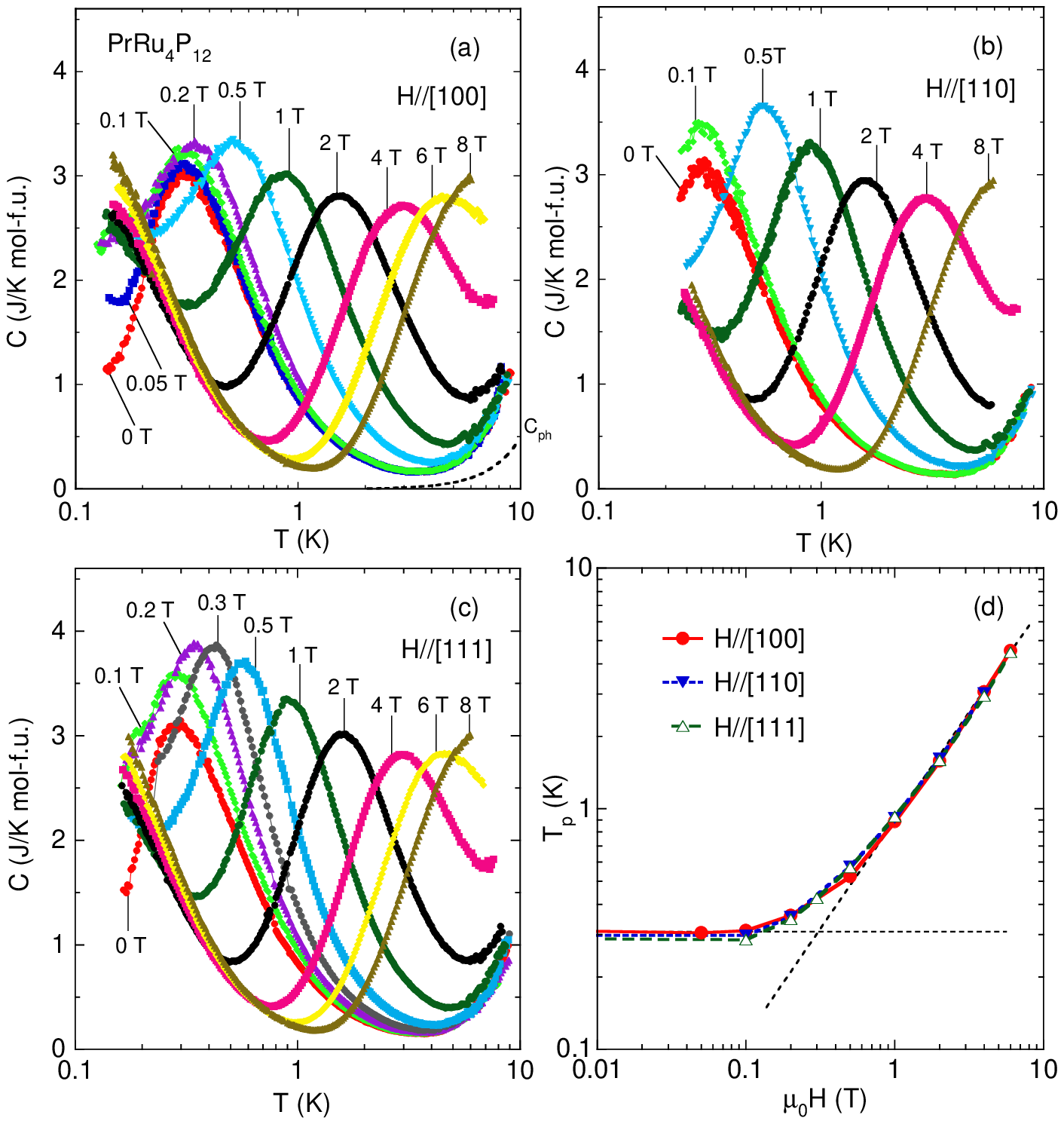} %14cm
\end{center}
\caption{(Color online) Specific heat $C(T,H)$ of PrRu$_4$P$_{12}$ for (a) $H \parallel [100]$, (b) $H \parallel [110]$, and (c) $H \parallel [111]$. The broken curve depicted in (a) represents the phonon part $C_{\rm ph}$ estimated from the $C(T)$ data of LaRu$_4$P$_{12}$. (d) The field dependence of the peak temperature $T_{\rm p}$. The dashed lines represent the low-$H$ and high-$H$ asymptotic behaviors.}
\label{fig:ct}
\end{fullfigure}

Figure~\ref{fig:ct} shows the specific heat $C$ for magnetic fields 
applied along the three principal directions.
The most significant finding is that a Schottky-type 
peak structure appears in zero field at $T_{\rm p}=0.30$ K and it 
shifts to higher temperatures with increasing field for all the 
field directions. 
It is obvious that this peak structure is caused by the thermal 
excitations in the Pr2 ions from the following reasons.
The phonon contribution $C_{\rm ph}=\beta T^3$ 
with $\beta=0.50$ mJ/K$^4$mol-f.u. determined for LaRu$_4$P$_{12}$ 
is negligibly small (see the dashed line in Fig.~\ref{fig:ct}).
Considering the extremely low carrier density in this $T$ 
range ($T \ll T_{\rm co}$)~\cite{Saha2009}, the conduction-electron 
contribution $C_{\rm e}=\gamma T$, where $\gamma$ must be two or 
three orders of magnitude smaller than 44.4 mJ/K$^2$mol-f.u. of 
LaRu$_4$P$_{12}$, should also be negligibly small. 
For the Pr1 ions, since the CEF ground state is $\Gamma_1$ 
singlet and the 1st excited state $\Gamma_4^{(2)}$ (triplet)  is well 
separated with the excitation energy of 94 K~\cite{IwasaPRB2005}, 
the Pr1-ion contributions are invisibly small in the present $T$ 
range, except that enhanced nuclear contribution appears 
as $\sim 1/T^2$ at low temperatures in applied fields 
(see Fig.~\ref{fig:Cmodel}).
As we demonstrate below, the peak structure can be accounted 
for as the thermal excitations in the $4f$-electron-nuclear 
hyperfine-coupled multiplets of Pr2 ions. 
The involvement of the Pr2 nuclear degrees of freedom in this peak 
is evidenced by the fact that the estimated entropy release from the Pr2 ions in the temperature range of 
0.13 to 4 K amounts to 9.18 J/Kmol-Pr2-ion, which exceeds already $R\ln 3$ expected from the $\Gamma_4^{(2)}$ triplet.

The field dependence of the peak temperature $T_{\rm p}$ plotted 
in Fig.~\ref{fig:ct}(d) indicates that there is a crossover 
at $\mu_0 H^* \sim$0.3 T from a low-$H$ regime with 
field-insensitive $T_{\rm p}$ to a high-$H$ regime with 
field-sensitive $T_{\rm p}$, as depicted by the two dashed lines.
In the latter regime, the slope of the linear behavior 
is estimated to be $T_{\rm p}/(\mu_0 H)=0.76 \pm 0.02 $ K/T 
in $\mu_0 H=2 \sim 6$ T.~\cite{SekineCpPB2000}
As shown in Fig.~\ref{fig:ct}(d), the magnetic anisotropy 
in the $T_{\rm p}$ vs $H$ is weak; it is only noticeable 
in the crossover field region ($H \sim H^*$).
In contrast, the anisotropy in the peak height $C(T_{\rm p})$ 
is more pronounced especially in the crossover field 
region: {\it e.g.}, 
$C(T_{\rm p}, H \parallel[111])/C(T_{\rm p}, H \parallel[100])=1.17$ 
in 0.2 T.

\begin{figure}[t] % 
\begin{center}
\includegraphics[width=14cm]{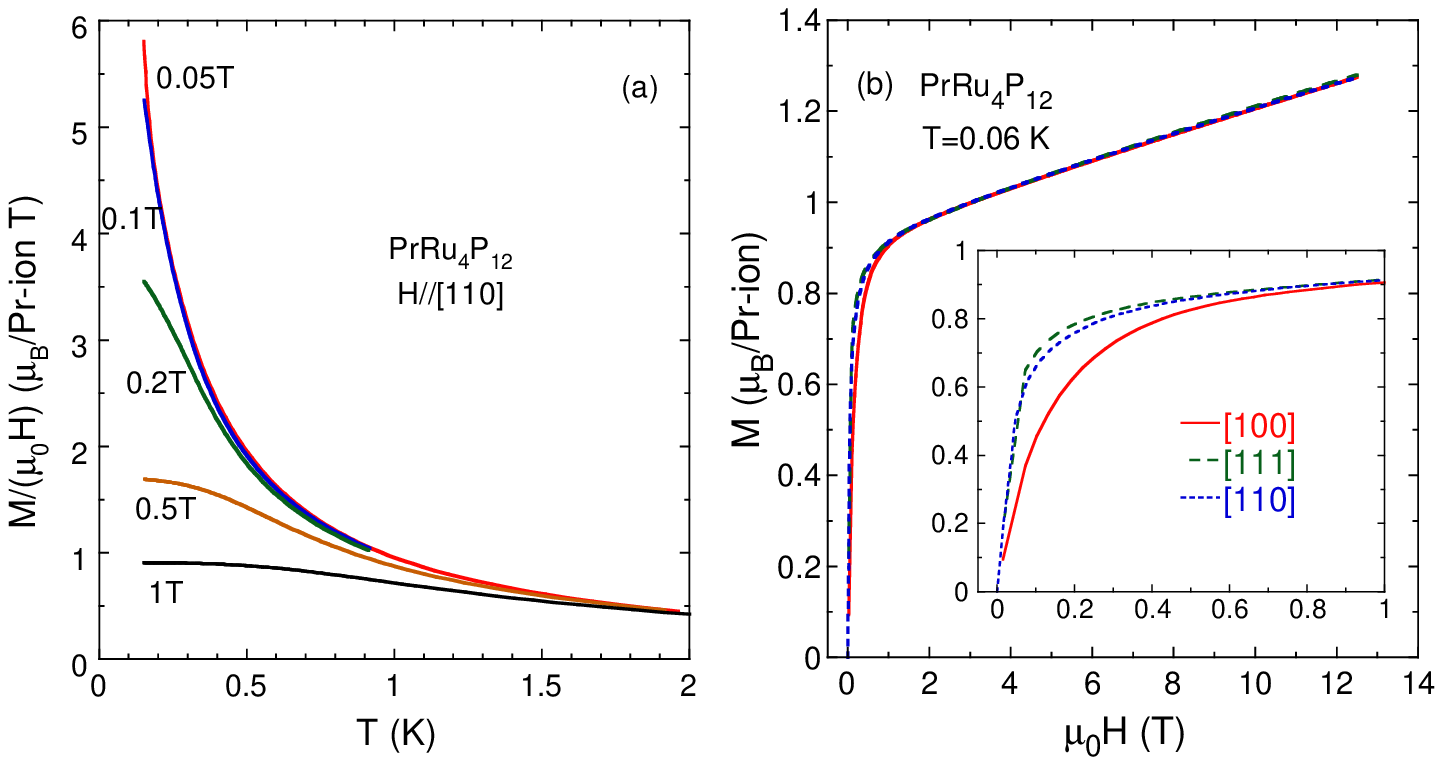}
\end{center}
\caption{(Color online) (a) The temperature dependence of magnetization 
$M(T)/(\mu_0 H)$ of PrRu$_4$P$_{12}$ measured for $H\parallel [110]$. 
(b) Isothermal magnetization curves measured at 0.06 K for the 
three principal directions [100], [110], and [111] (data 
originally reported in ref.~\citen{IwasaPRB2005}). 
An expanded view for $\mu_0 H <1$ T is shown in the inset . }
\label{fig:M}
\end{figure}

Figure~\ref{fig:M}(a) shows the temperature dependence of magnetization 
measured for several fixed fields along the [110] direction.
For  $\mu_0 H <0.2$ T, $M/(\mu_0 H)$ shows a Curie-type 
divergence at low temperatures as expected for the 
magnetically degenerate $\Gamma_4^{(2)}$ triplet ground state of Pr2 ions.
With increasing $H$ the divergence becomes weaker, and low-$T$ 
saturation becomes pronounced for $\mu_0 H >0.2$ T.
Isothermal magnetization curves at 0.06 K are 
shown in Fig.~\ref{fig:M}(b). Above $\sim 2$ T, $M(H)$ increases almost 
linearly up to at least $\sim 12.5$ T.
In the high-$H$ region, it has been reported that $M(H)$ data agree well 
with those calculated using the CEF level schemes of Pr1 and Pr2 ions 
determined by inelastic neutron scattering (INS)~\cite{IwasaPRB2005}.
In contrast, the CEF model fails to reproduce the $M(H)$ data below 1 T.
As shown in the inset, $M(H)$ shows a significant rounding and 
a downward curvature remains up to $\sim 1$ T.
This anomalous rounding, which cannot be explained with the Brillouin function 
for the $\Gamma_4^{(2)}$ triplet ground state of Pr2 ion, 
will be discussed below (see Fig.~\ref{fig:Mmodel}).
In this field region, magnetic anisotropy is noticeable with 
$M(H \parallel [111]) \gtrsim M(H \parallel [110]) > M(H \parallel [100])$.
This anisotropy seems to be thermodynamically consistent 
with the anisotropic peak height 
$C(T_{\rm p}, H \parallel [111]) \gtrsim C(T_{\rm p}, 
H \parallel [110]) > C(T_{\rm p}, H \parallel [100])$ 
(see Fig.~\ref{fig:ct}), i.e., specific heat is relatively insensitive to fields when applied along the hard magnetization axis [100].

\section{Model Calculation} % --------------

Now we study the microscopic origin of the observed 
low-temperature behaviors of PrRu$_4$P$_{12}$.  Let us note again that there is no experimental 
indication of long range ordering within the temperature range explored,  though the CEF ground state on one of the sublattice remains triply degenerate. This fact is in accord with the theoretical study showing that 
the effective magnetic interaction among the triplets in the charge 
order phase is  extremely small.\cite{ShiinaShibaJPSJ2010}   
On the other hand, it is known that the hyperfine interaction 
between nuclear spin and $4f$ electron in Pr ion can influence 
on the thermodynamic quantities in the temperature 
region of the order of 0.1 K. Therefore, we analyze 
here possible consequences of the hyperfine interaction 
at the single Pr site. 

In a Pr ion, $^{141}$Pr nucleus (the natural abundance of 100\%) has 
a nuclear spin $I=5/2$ so that 
we start from the following simple and well-established 
Hamiltonian~\cite{Murao1971,Ishii2004,Bleaney1973} to deal 
with each Pr ion (Pr1 and Pr2);
\begin{equation}
 \label{eq:Hamil} {\mathcal H} = {\mathcal H}_{\rm CEF}
+A {\mib I} \cdot {\mib J}-(-g_J \mu_{\rm B} {\mib J}
+g_{\rm N} \mu_{\rm N} {\mib I}) \cdot \mu_0 {\mib H}.
\end{equation}
The first term corresponds to the CEF Hamiltonian for the cubic 
$T_h$ site symmetry,
\begin{equation}
 \label{eq:HCEF} {\mathcal H}_{\rm CEF}=A_4 ({\boldmath O}_{4}^0
+5 {\boldmath O}_{4}^4)+A_6^c ({\boldmath O}_{6}^0
-21 {\boldmath O}_{6}^4)+A_6^t ({\boldmath O}_{6}^2-{\boldmath O}_{6}^6), 
\end{equation}
where ${\boldmath O}_{m}^n$'s are  Stevens' operator 
equivalents~\cite{Takegahara2001}. For the CEF parameters 
${A_4, A_6^c, A_6^t}$, we use the set of values determined 
at 5 K by INS~\cite{IwasaPRB2005}. The second term in 
eq.~(\ref{eq:Hamil}) represents the hyperfine interaction 
of a Pr ion and we use the coupling constant $A=+0.052$ K, 
given by theoretical calculations~\cite{KondoHC1961,BleaneyHC1963} 
and confirmed later 
by thermodynamical measurements for PrFe$_4$P$_{12}$ and 
PrOs$_4$Sb$_{12}$~\cite{AokiPFPPRB2002,AokiJPSJ2002,SakakibaraJPSJ2008}.
We have neglected the quadrupolar hyperfine interaction, 
whose energy scale is calculated to be less than 
$\mid PI(2I-1)/3 \mid \sim 5 \times 10^{-4} $ K 
(see ref.~\citen{BleaneyHC1963} for the quadrupole coupling constant $P$).
The third term in eq.~(\ref{eq:Hamil}) represents the Zeeman energy 
of the magnetic moments due to the 4$f$-electron 
$-g_J \mu_{\rm B} {\mib J}$ and due to the nuclear spin 
$g_{\rm N} \mu_{\rm N} {\mib I}$, where $g_J=4/5$, 
$\mu_{\rm B}$, $g_{\rm N}=+1.72$, and $\mu_{\rm N}$ are 
the Land\'e $g$-factor, Bohr magneton, the nuclear $g$-factor, 
and the nuclear magneton, respectively. 
Note that there are no fitting parameters in eq.~(\ref{eq:Hamil}).

\begin{figure}[t] % 
\begin{center}
\includegraphics[width=7cm]{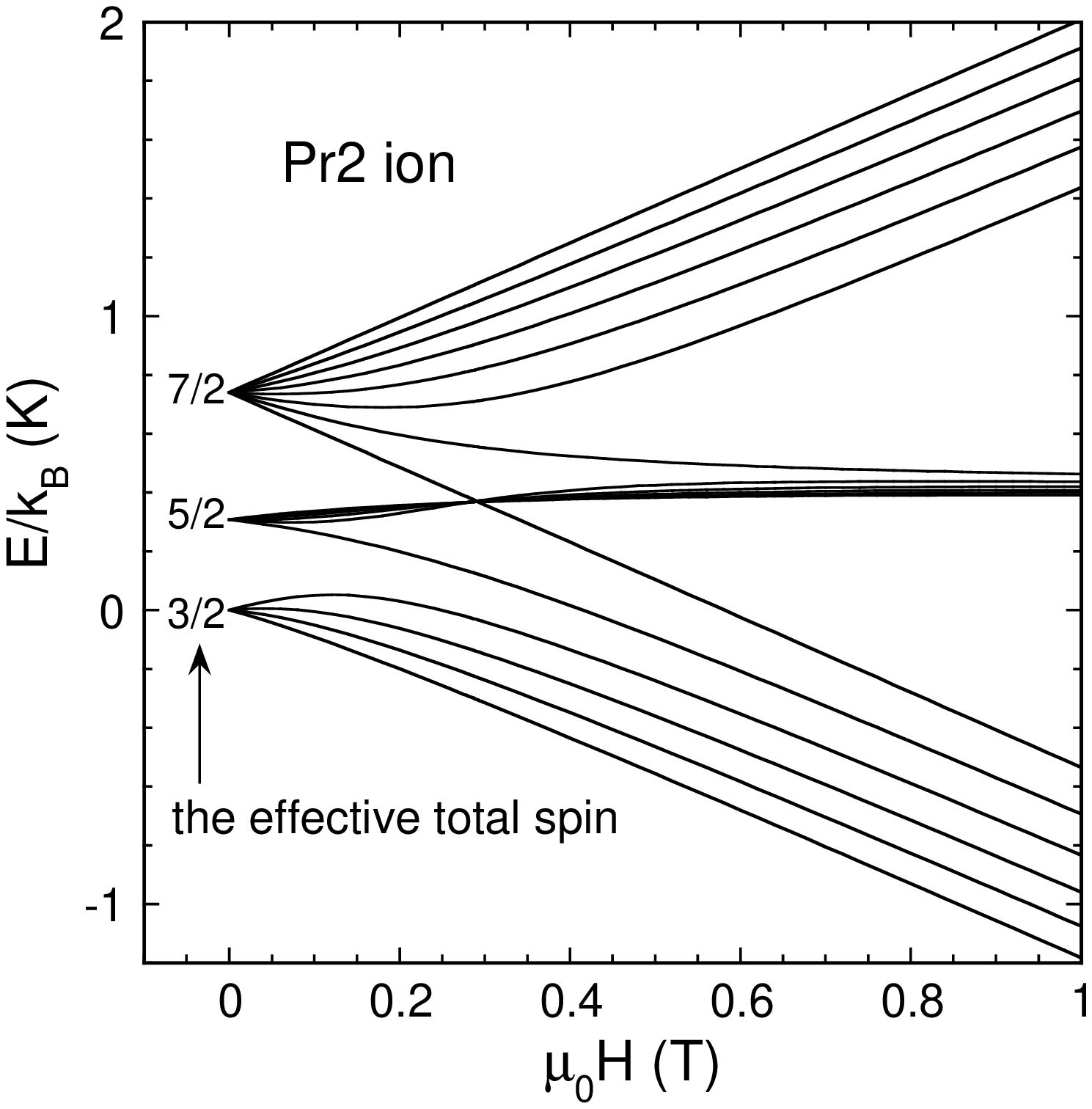}
\end{center}
\caption{Calculated energy level scheme of the 4$f$-electron-nuclear 
hyperfine-coupled multiplets for the Pr2 $\Gamma_4^{(2)}$ triplet 
as a function of applied magnetic field. No visible field direction dependence 
appears in the low $H$ region.}
\label{fig:EH}
\end{figure}

The energy level scheme in applied fields is obtained by diagonalizing the Hamiltonian. 
Figure~\ref{fig:EH} shows the low-energy region for Pr2 ion. 
At zero field the $\Gamma_4^{(2)}$ triplet of Pr2 
ion is reconstructed into three multiplets through the hyperfine 
coupling with the nuclear spin. This feature is well understood 
by using the pseudo-spin representation of the interaction 
as described in the Appendix. 
In contrast, the Pr1 ion has the energetically well-separated $\Gamma_1$ 
singlet ground state so that a sextet nuclear state (corresponding to $I_z=+5/2 \sim -5/2$) is left in the low-energy region. In applied fields, the Van-Vleck magnetism of 
4$f$-electrons leads to the enhanced nuclear magnetism on Pr1 ion 
through the hyperfine coupling.
Note that, in $\mu_0 H < 1$ T (as shown in Fig.~\ref{fig:EH}), 
the field direction dependence of the energy level scheme is 
hardly noticeable.

\begin{figure}[t] % 
\begin{center}
\includegraphics[width=7cm]{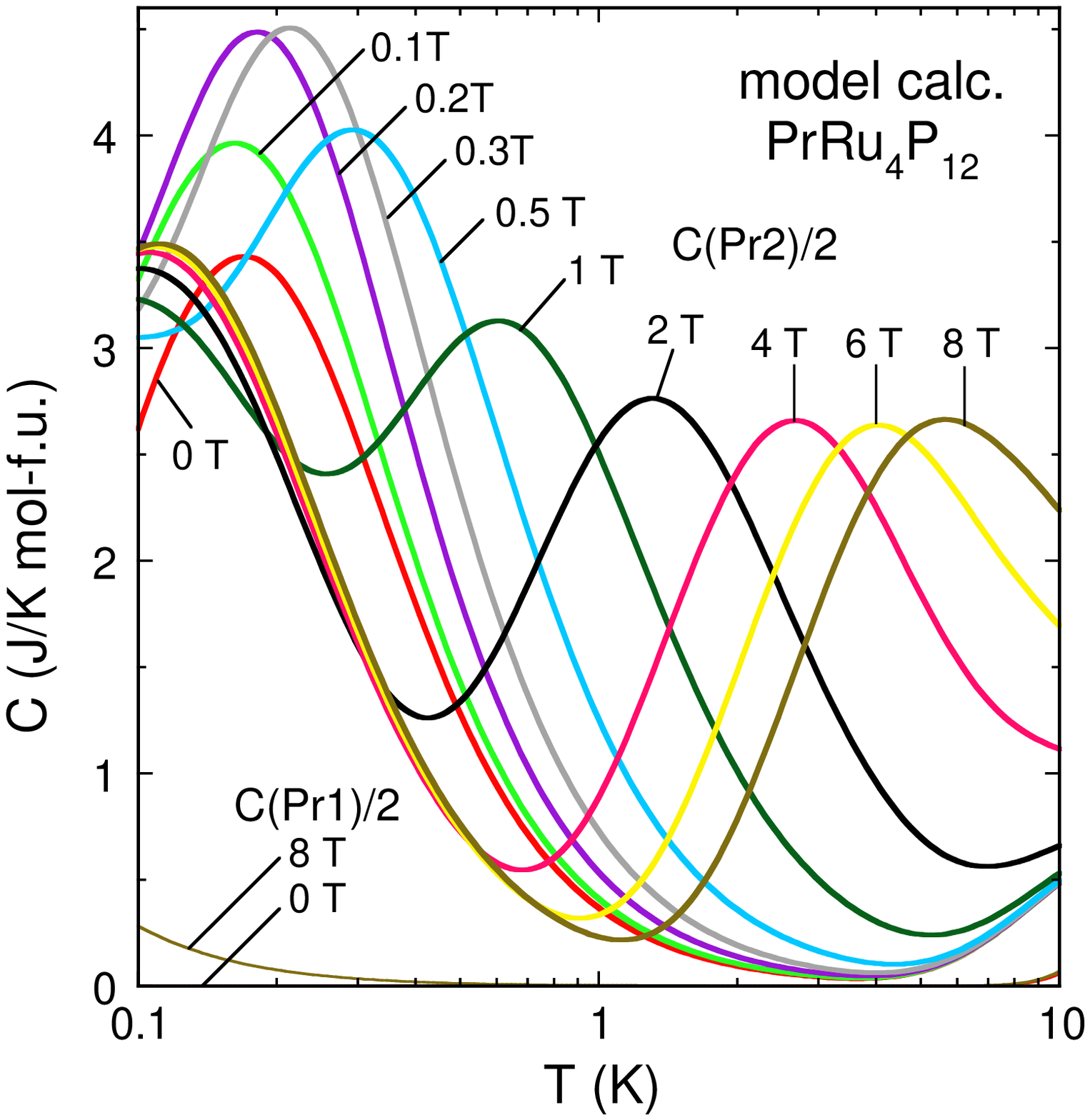}
\end{center}
\caption{(Color online) Model calculation of the specific heat $C(T, H)$ for PrRu$_4$P$_{12}$. Contributions from Pr2 ion ($C$(Pr2)/2) and from Pr1 ion ($C$(Pr1)/2) are shown separately. No noticeable magnetic anisotropy is present in the $T$ and $H$ regions shown here. $C$(Pr1)/2 (shown by thin lines for $\mu_0H=0$ and 8 T) has a very small contribution. 
Contributions from Ru and P nuclei are negligibly small\cite{RuPNuc}.}
\label{fig:Cmodel}
\end{figure}

The $T$ and $H$ dependences of the specific heat are calculated separately 
for Pr2-ion ($C$(Pr2)) and Pr1-ion ($C$(Pr1)) 
(see Fig.~\ref{fig:Cmodel}). 
The experimental data of $C$ should be compared with 
$C$(Pr2)/2+$C$(Pr1)/2 since one formula unit of PrRu$_4$P$_{12}$ 
contains 0.5$\times$Pr2-ion and 0.5$\times$Pr1-ion.
At zero field, the Pr1 contribution $C$(Pr1)/2 is almost zero 
(slight increase visible around 10 K is due to the 
$\Gamma_1-\Gamma_4^{(2)}$ excitation). At  8 T 
it is still small enough compared to $C$(Pr2)/2, although the 
hyperfine-enhanced nuclear contribution shows up as $C$(Pr1)$\propto (AM({\rm Pr1})/T)^2$ below $\sim 0.5$ K.

It is evident that the observed peak structure for $H=0$ is 
reasonably well reproduced by $C$(Pr2)/2. From the energy level scheme shown 
in Fig.~\ref{fig:EH}, it is clear that the peak structure originates 
from the thermal excitations among the three multiplets.
Note that if the value of $A$ were zero then the three-multiplet 
structure would collapse and the $C(T)$ peak structure would 
disappear.
Since the three multiplets have fixed ratios of the energy 
separation ($-7/2:-1:+5/2$) and the degeneracy ($4:6:8$) 
as shown in the Appendix, the peak height $C(T_{\rm p})=3.43$ J/Kmol-f.u. 
is a universal $A$-independent constant.
The observed value 3.03 J/Kmol-f.u. is close to this theoretical expectation.
The increase in $C$(Pr2)/2 above $\sim 5$ K is due to the thermal 
excitations to the 1st excited $\Gamma_1$ lying at 
36 K.~\cite{IwasaPRB2005}
In high fields, the peak shifts to higher temperatures and 
the height approaches 2.65(=5.30/2) J/Kmol-f.u., which is the expected 
value for a triplet with an equal energy separation.

\begin{figure}[t] % 
\begin{center}
\includegraphics[width=14cm]{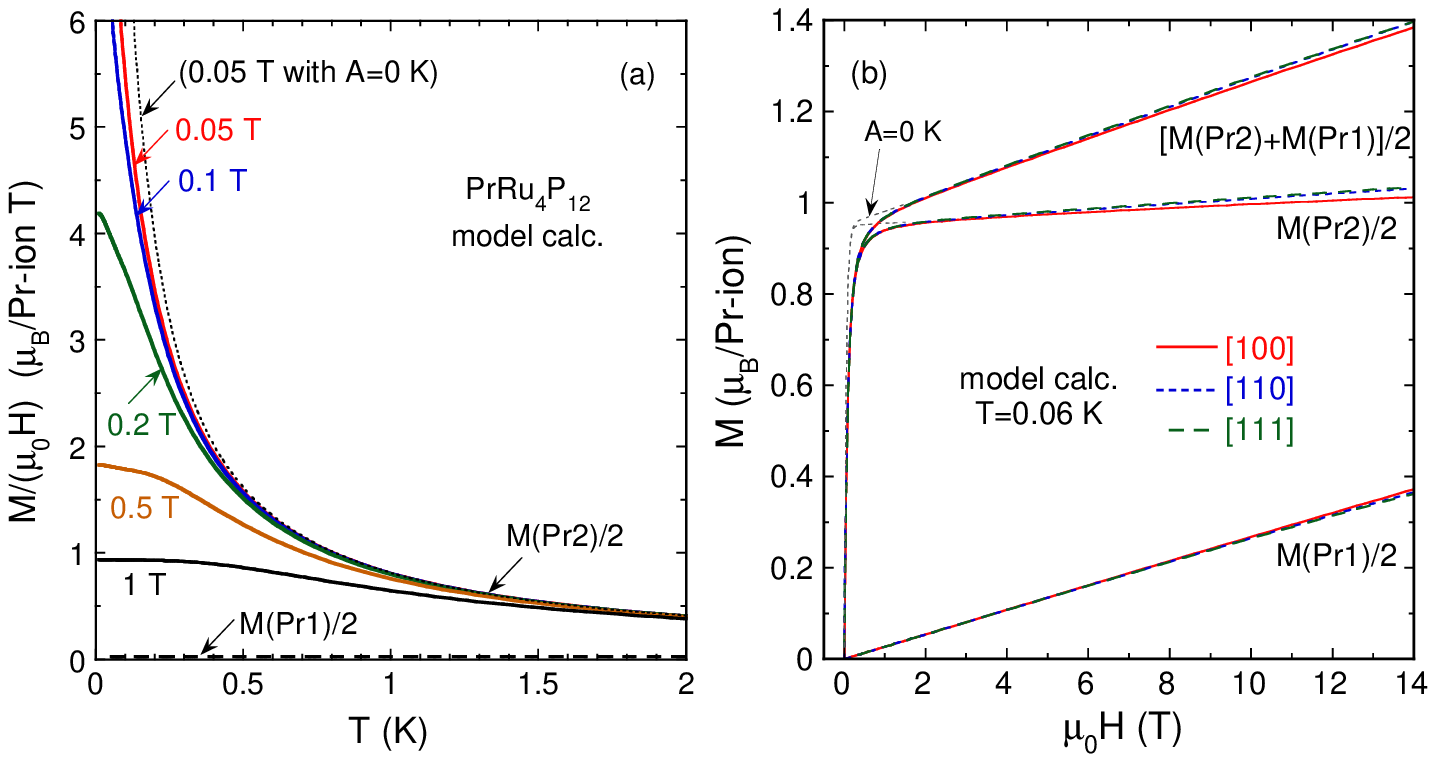}
\end{center}
\caption{(Color online) Model calculation of the magnetization $M(T,H)$ for  
PrRu$_4$P$_{12}$. $M$(Pr1)/2 and $M$(Pr2)/2 represent the 
contributions from Pr1 and Pr2 ions, respectively. (a) The $T$ 
dependences of $M$ for several fields. $M$(Pr2)/2 at 0.05 T 
for a fictitious case with $A=0$ is drawn with  thin broken lines. 
(b) Isothermal magnetization curves at 0.06 K for  [100], [110], and [111]. At  high fields, 
the small magnetic anisotropy with 
$M(H \parallel [111]) \gtrsim M(H \parallel [110]) 
> M(H \parallel [100])$ 
is due to $M$(Pr2)/2 ($M$(Pr1)/2 has the opposite anisotropy 
with a smaller magnitude). The calculation for $A=0$ is shown with thin broken lines below 2 T.}
\label{fig:Mmodel}
\end{figure}

Concerning the magnetization, as shown in Fig.~\ref{fig:Mmodel},  the Pr2 contribution $M$(Pr2)/2 dominates over the Pr1 
contribution $M$(Pr1)/2.  $M(T)/(\mu_0H)$ for $\mu_0H \le 1$ T 
shown in Fig.~\ref{fig:Mmodel}(a) agree quite well with the 
observed data shown in Fig.~\ref{fig:M}.
A  fictitious case corresponding to no hyperfine coupling ($A=0$) is also drawn for $M$(Pr2)/2 at 0.05 T. 
Comparing the two curves, we find  that a partial suppression 
in the low-$T$ Curie-type divergence is caused by the hyperfine 
coupling. 
Our model calculation of magnetization curves at 0.06 K shown in 
Fig.~\ref{fig:Mmodel}(b) also reproduces well the measured data 
shown in Fig.~\ref{fig:M}(b).
At high fields, the small magnetic anisotropy 
with $M(H \parallel [111]) \gtrsim M(H \parallel [110]) 
> M(H \parallel [100])$ in the calculated $M$ agrees with that 
in the measured data, confirming the reported CEF parameters 
of the Pr ions~\cite{IwasaPRB2005}.
Below 2 T, the observed anomalous rounding in the $M(H)$ curves 
is nicely reproduced by the calculation. 
Comparing with the calculation for $A=0$, we conclude  that this 
rounding is due to the hyperfine coupling.

\section{Discussions}

We have shown that the overall behaviors of the observed low-temperature 
anomalies, i.e., the peak structure in $C(T)$ and the rounding 
in the low-field $M(H)$ curves, provide an evidence for the formation 
of the 4$f$-electron-nuclear hyperfine-coupled multiplets on Pr2 
ions. 
Let us note that the hyperfine-coupled multiplet formation has been observed on the periodic 
Pr lattice.
The $\Gamma_4^{(2)}$ triplet of Pr ion has active magnetic dipole and 
electric quadrupole moments. 
Usually, inter-ion interactions of those moments in compounds 
tend to drive them into ordering at low temperatures.
Actually, in PrB$_6$~\cite{LoewenhauptZP1986}, 
PrOs$_4$As$_{12}$~\cite{MaplePNAS2006}, and 
PrCu$_4$Au~\cite{ZhangJPCM2009}, all of which have a triplet 
ground state, antiferromagnetic orderings set in at 7, 2.3, and 2.5 K, 
respectively.
In this respect the charge-ordered phase in PrRu$_4$P$_{12}$ 
is a rare system, where the hyperfine-coupled multiplet lattice of the triplet 
Pr ions is observed without showing any phase transition  
down to 60 mK. This remarkable feature is produced by the charge 
ordering below $T_{\rm co}$, giving rise to the large Pr2-Pr2 
distance (equal to the cubic lattice constant 
$a$=8.0420(9) \AA~\cite{Jeitschko1977}) and the disappearance 
of a large amount of conducting electron carriers, both of 
which weaken inter-Pr-ion RKKY-type exchange interactions.
Furthermore, by the characteristic electronic structure, 
where the interactions among the 4f-electrons of Pr ions are 
mediated by mixing with the $a_u$ molecular orbital on P$_{12}$, 
magnetic interactions are suppressed as demonstrated 
by a microscopic calculation~\cite{ShiinaShibaJPSJ2010}.

Although the present model successfully accounted for the overall behaviors of the observed low-temperature 
anomalies, there still remain following discrepancies between the measurements 
and the calculations:  (i) the observed $T_{\rm p}=0.30$ K is 
higher than the calculated value of 0.17 K and (ii) the anomalous 
magnetic anisotropies appear in $C$ and $M$ in the crossover 
field region of $H \sim H^*$.

Within the model (\ref{eq:Hamil}), in order to account for the observed $T_{\rm p}=0.30$ K at zero field, a larger value of $A=+0.091$ K would be needed. 
However, this leads to an inconsistency. 
The value of $A$ can be accurately estimated with the $C(T)$ data 
at  $\mu_0H=8$ T.  Below 1 K, $M$(Pr2)$=g_J \langle J \rangle_{\rm Pr2}$ 
as well as $M$(Pr1)$=g_J \langle J \rangle_{\rm Pr1}$ is saturated 
and therefore the low-$T$ increase in $C(T)$ is dominated by the 
Zeeman splitting of the nuclei, which feel the effective 
hyperfine-enhanced magnetic field of 
$\mu_0H-A \langle J \rangle/g_{\rm N} \mu_{\rm N}$ for each Pr ion.
From the fitting to the $C(T)$ data shown in Fig.~\ref{fig:ct}, 
we obtain $A=+0.050(2)$ K for PrRu$_4$P$_{12}$.
The values for $A$ in metallic 
PrFe$_4$P$_{12}$~\cite{AokiPFPPRB2002} and 
PrOs$_4$Sb$_{12}$~\cite{AokiJPSJ2002,SakakibaraJPSJ2008} 
and extremely-low-carrier PrRu$_4$P$_{12}$ are remarkably close to each other.
This means that the intra-ion $4f$-electron-nuclear hyperfine coupling is not 
affected by the background electronic band structure in materials. 

There might be other interactions missing in eq.~(\ref{eq:Hamil}) that possibly yields 
additional energy splittings in the multiplets, whereby resulting in the higher 
value of $T_{\rm p}$.
The magnetic dipole and electrical quadrupole moments of 
the $\Gamma_4^{(2)}$ triplet can play a role in such interactions.
However, it is difficult to explore this scenario further 
since no observations have been reported to date 
on time-reversal symmetry breaking and Pr2-site local symmetry lowering.

As another possible explanation, the discrepancies may be caused 
by a many body effect associated with the $c$-$f$ hybridizations.
The 4$f$-electron-nuclear hyperfine-coupled multiplets have 
composite multipole moments and these internal degrees of freedom 
can be coupled to the slightly remaining conduction electrons 
in the charge-ordered state. The anomalously $T$- and $H$-dependent 
transport properties observed below $\sim 10$ K~\cite{Saha2009} 
might be reflecting this effect.
This interesting possibility should be investigated in future studies. 

Finally we note that the antiferromagnetic hyperfine coupling between the nuclear spin $I=5/2$ and $\Gamma_4^{(2)}$ triplet leaves the effective total spin of $F'=3/2$ at low temperatures ($T\ll A$).  Needless to say, this finite degree of freedom should be lifted at very low temperatures via inter-ion interactions. 

\section{Summary} % ---------------------

Specific heat and magnetization 
measurements have been carried out to investigate the low-temperature 
properties of the Pr2-ion CEF {\it triplet} ground state 
of the charge order phase in PrRu$_4$P$_{12}$.
It has been revealed that the specific heat shows a  Schottky-type broad 
peak structure at $T_{\rm p}=0.30$ K at zero field and the magnetization 
curve at 0.06 K shows a remarkable rounding below 1 T.
We have demonstrated that these anomalous behaviors can be well 
explained by taking into account the hyperfine coupling of 
Pr2 ions, thus showing  that the {\it 4$f$-electron-nuclear 
hyperfine-coupled multiplets} are formed in PrRu$_4$P$_{12}$. 
A comparison with the model calculation shows that there still 
remain some discrepancies to be understood, i.e., the higher 
value of $T_{\rm p}$ in zero field and the small low-$H$ magnetic 
anisotropies as demonstrated in the $C(T)$ and $M(H)$ data.
Since 4$f$-electron-nuclear hyperfine-coupled multiplets have high degeneracies, 
the associated multipolar degrees of freedom might be relevant 
to the appearance of the discrepancy, possibly by a many body 
effect through the $c$-$f$ hybridizations.

\section*{Acknowledgments}

We thank H. Harima, H. Ishii, K. Iwasa, T. Mito, and C. Sekine for valuable discussions.
This work was supported by the Grant-in-Aid for 
Scientific Research on Priority Area "Skutterudite" (15072206) and "Ubiquitous" (20045015),
and on Innovative Areas "Heavy Electrons" (20102007, 21102520) 
of MEXT and (C: 20540359, 21540368) and (B: 20340094) of JSPS, Japan.

\appendix

\section{ Pseudo-spin Description of Hyperfine Interaction }

In eq. (1), the CEF Hamiltonian ${\mathcal H}_{\rm CEF}$ leaves 
the triple degeneracy at the Pr2 site. Here we study 
how this degeneracy is affected by the hyperfine interaction 
$A {\mib I} \cdot {\mib J}$ using 
the pseudo-spin representation. 
The triplet $\Gamma_4^{(2)}$ at the Pr2 site is given
as follows  
\begin{align}
|\Gamma_4^{(2)}(j)\rangle 
=\sqrt{1-d^2} |\Gamma_5(j)\rangle + d |\Gamma_4(j))\rangle \ ,
\end{align}
where $j$ is an index to specify  each triplet ($j=+, 0, -$).
$|\Gamma_5(j)\rangle$ and $|\Gamma_4(j)\rangle$ are two sets of triplet
wave functions for the $O_{h}$ group: 
\begin{align}
|\Gamma_5(\pm)\rangle 
& =\pm\sqrt{{7 \over 8}}|\pm 3\rangle \mp
 \sqrt{{1 \over 8}} | \mp 1\rangle \ ,
\\
|\Gamma_5(0)\rangle
&=\sqrt{{1 \over 2}}\big( | 2 \rangle -  | -2
 \rangle \big) \ ,
\\
|\Gamma_4(\pm)\rangle
&=\mp \sqrt{{1 \over 8}}|\mp 3\rangle \mp
 \sqrt{{7 \over 8}} | \pm 1\rangle \ ,
\\
|\Gamma_4(0)\rangle
&=\sqrt{{1 \over 2}}\big( | 4 \rangle -  | -4
 \rangle \big) \ .
\end{align}
The coefficient $d$ in eq. (A.1) is a parameter of mixing due to $T_h$, 
which is known to be small in PrRu$_4$P$_{12}$.

We introduce the $S=1$ pseudo-spin operator ${\mib \tau}$ and 
regard the $\Gamma_4^{(2)}$ triplet with $j=\pm, 0$ as the 
pseudo-spin states corresponding to  $\tau_z=\pm1, 0$. Then, the original 
dipole operator is represented as 
$ {\mib J}= c {\mib \tau}  $ with $ c = 5/2 - 2d^2 $. 
Therefore, the hyperfine interaction $A {\mib I} \cdot {\mib J}$ 
can be replaced by $\lambda {\mib I} \cdot {\mib \tau}$ 
with an effective coupling constant $\lambda=c A$. 
Then, it is clear that the degenerate $18(=3 \times 6)$ states 
$\mid \tau_z, I_z \rangle$ are split into three multiplets.
Introducing an effective total spin ${\mib F}'={\mib \tau}+{\mib I}$, 
the three multiplets can be labeled as $F'=3/2, 5/2$, and 7/2.
It is interesting to note that the half-integer spin multiplets 
are formed from the integer-spin $4f$-electron states of non-Kramers 
Pr ion through the hyperfine coupling with the half-integer-spin 
$^{141}$Pr nucleus. The characteristic features of each multiplet 
are listed in Table~\ref{table:1}.

\begin{table}[h]
\caption{Characteristic features of the three multiplets formed 
by the hyperfine coupling on Pr2 ion. The effective total spin $F'$, 
the energy $E_i$, the effective $g$-factor $g^*$, and the magnetic 
moment $-g^* \mu_{\rm B} F'$ are listed. 
}
\label{table:1}
\begin{center}
\begin{tabular}{@{\hspace{\tabcolsep}\extracolsep{\fill}}ccccccc}
\hline
$F'$ (degeneracy) & energy $E_i$ & $E_i-E_0$ & $g^*$ & $-g^*F'$ \\
           &     &  (K) &   &  ($\mu_{\rm B}$)  \cr
\hline
7/2 (8) & $E_2=+5/2 \lambda$ &  +0.74 &  +0.541 &  -1.894 \cr
5/2 (6) & $E_1=- \lambda$    &  +0.31 &  +0.216 &  -0.540 \cr
3/2 (4) & $E_0=-7/2 \lambda$ &  0     &  -0.760 &  +1.140 \cr
\hline
\end{tabular}
\end {center}
\end{table}

The magnetic moment of each multiplet can be expressed 
as $-g^* \mu_{\rm B} {\mib F}'$, where the effective $g$-factor 
$g^*$ has a specific value for each multiplet as listed 
in Table~\ref{table:1}. When a weak magnetic field is applied, 
each multiplet shows the Zeeman splitting (see Fig.~\ref{fig:EH}) 
with the $g^*$ value, independent on the field direction.
With further increasing the field, the Zeeman energy becomes dominant 
over the hyperfine coupling energy and the 18 states are 
asymptotically separated into three sextets, i.e., 
$\mid +1, I_z \rangle$, $\mid 0, I_z \rangle$, 
and $\mid -1, I_z \rangle$ with $I_z=+5/2 \sim -5/2$ 
in high fields (so called the Paschen-Back effect).
For $\mid \pm 1, I_z \rangle$, since the $4f$-electron 
polarization $\langle {\mib J} \rangle$ provides strong effective 
field $-A \langle {\mib J} \rangle/g_{\rm N} \mu_{\rm N}$ 
on the Pr nucleus, both sextet splits into six equally-spaced lines, 
in contrast to the degenerate sextet for $\mid 0, I_z \rangle$.
The crossover field between the two field regimes appears 
to be $\sim 0.3$ T, which is consistent with $\mu_0 H^* \sim$0.3 T 
in Fig.~\ref{fig:ct}(d).

\end{document}